\documentclass[fleqn,usenatbib]{mnras}

\usepackage{newtxtext,newtxmath}
\usepackage{subfig}

\usepackage[T1]{fontenc}
\usepackage{ae,aecompl}
\usepackage[acronym]{glossaries}
\usepackage{hyperref}
\usepackage{comment}
\usepackage{newtxtext,newtxmath}

\usepackage{graphicx}	
\usepackage{amsmath}	
\usepackage{epsfig}
\usepackage{xcolor}

\newcommand{\hi}{\ion{H}{I}}
\def\kk{{\mathbfit{k}}}
\usepackage{tikz}
\def\checkmark{\tikz\fill[scale=0.4](0,.35) -- (.25,0) -- (1,.7) -- (.25,.15) -- cycle;} 

\title[PS using ANNs]{Extracting the 21-cm Power Spectrum and the reionization parameters from mock datasets using Artificial Neural Networks}

\author[M.Choudhury et al]{
Madhurima Choudhury$^{1}$\thanks{E-mail: madhurimachoudhury811@gmail.com},
{Abhirup Datta$^{1}$,~}{Suman Majumdar$^{1,2}$}
\\
$^{1}$Department of Astronomy, Astrophysics Indian Institute of Technology Indore, India\\
$^2$Department of Physics, Blackett Laboratory, Imperial College, London SW7 2AZ, UK
}

\date{Accepted XXX. Received YYY; in original form ZZZ}

\pubyear{2021}
\begin{document}
\label{firstpage}
\pagerange{\pageref{firstpage}--\pageref{lastpage}}
\maketitle

\begin{abstract}
Detection of the \hi~ 21-cm power spectrum is one of the key science drivers of several ongoing and upcoming low-frequency radio interferometers. However, the major challenge in such observations come from bright foregrounds, whose accurate removal or avoidance is key to the success of these experiments. In this work, we demonstrate the use of artificial neural networks (ANNs) to extract the \hi~ 21-cm power spectrum from synthetic datasets and extract the reionization parameters from the \hi~ 21-cm power spectrum. For the first time, using a suite of simulations, we present an ANN based framework capable of extracting the \hi~ signal power spectrum directly from the total observed sky power spectrum (which contains the 21-cm signal, along with the foregrounds and effects of the instrument). To achieve this, we have used a combination of two separate neural networks sequentially. As the first step, \texttt{ANN1} predicts the 21-cm power spectrum directly from foreground corrupted synthetic datasets. In the second step, \texttt{ANN2} predicts the reionization parameters from the predicted \hi~ power spectra from \texttt{ANN1}. Our ANN-based framework is trained at a redshift of $9.01$, and for \kk-modes in the range, $\rm{0.17<\kk<0.37~Mpc^{-1}}$. We have tested the network's performance with mock datasets that include foregrounds and are corrupted with thermal noise, corresponding to $1080$ hrs of observations of the \textsc{ska-1 low} and \textsc{hera}. Using our ANN framework, we are able to recover the \hi~ power spectra with an accuracy of $\approx95-99\%$ for the different test sets. For the predicted astrophysical parameters, we have achieved an accuracy of $\approx~81-90\%$ and $\approx~50-60\%$ for the test sets corrupted with thermal noise corresponding to the \textsc{ska-1 low} and \textsc{hera}, respectively.


\end{abstract}



\section{Introduction}

The redshifted 21-cm line of neutral hydrogen is a sensitive probe to investigate the different phases of the evolution of our Universe. Observations of this 21-cm line will directly enable us to map the young Universe, over a range of cosmic times and give us deep insight into the morphology of the ionisation structures which were carved out by the first sources of light, as well as about the origin and evolution of these first generation sources. \citep{Morales_2010, Pritchard_2012, Furlanetto2016}.  The state of the intergalactic medium, right from the period when the first stars and galaxies were formed (Cosmic Dawn or CD) through the period when the Universe became completely ionized (Epoch of Reionization or EoR) and evolved to the Universe we see today, can be probed with the 21-cm line. The measurement of \hi~ 21-cm power spectrum using large interferometric arrays currently hold the greatest potential to observe the redshifted \hi~ 21-cm line \citep{Bharadwaj_2001, Bharadwaj_2005, Morales_2005, Zaldarriaga_2004} and probe the large-scale distribution of neutral hydrogen across a range of redshifts. Measurements of the \hi~ 21-cm power spectrum is a major goal of several ongoing and future experiments. Several radio interferometers such as the Giant Meterwave Radio Telescope (GMRT, \citet{Swarup_1991}); the Low Frequency Array (LOFAR, \citet{Harlem_2013});  
the Murchison Wide-field Array (MWA, \citet{Tingay_2013}); 
and the Donald C. Backer Precision Array to Probe the Epoch of Reionization (PAPER, \citet{Parsons_2010}) have carried out observations to help constrain the 21-cm power spectrum from the Epoch of Reionization. Future experiments like the Hydrogen Epoch of Reionization Array (HERA, \citet{DeBoer_2017}) and the Square Kilometer Array (SKA, \cite{Koopmans_2015})  also aim to measure
the EoR 21-cm power spectrum with much improved sensitivities promising to give a deeper insight into the physics of the evolution of the Universe. Experiments such as, the Experiment to Detect the Global EoR Signature (EDGES), \citep{Bowman_2018}; Shaped Antenna measurement of the background RAdio Spectrum (SARAS), \citep{singh2021}; the Large-Aperture Experiment to Detect the Dark Ages (LEDA), \citep{Greenhill_2012}; SCI-HI, \citep{Voytek_2014}; the Broadband Instrument for Global Hydrogen Reionisation Signal (BIGHORNS) \citep{Sokolowski_2015}, 
and the Cosmic Twilight Polarimeter, CTP \citep{Nahn_2018} aim to measure the sky averaged global 21-cm signature.

However, the detection of this faint \hi~ 21-cm signal is quite challenging. This is because it is buried in a sea of galactic and extragalactic foregrounds, which are several orders of magnitude brighter than the signal. In addition, the instrument response also varies with increased observation times. The Earth's ionosphere also further distorts the signal by introducing direction-dependent effects, making ground-based observations even more challenging. Several novel techniques have been explored to remove bright foregrounds from both interferometric as well as total power experiments. The 21-cm observations heavily rely on the accuracy of foreground removal and instruments with high sensitivity. The foreground sources comprise of diffuse Galactic synchrotron emission from our Galaxy \citep{Shaver_1999}, free-free emission from ionizing haloes \citep{Oh_2003}, synchrotron emission and radiation from comptonization processes from faint radio-loud quasars \citep{DiMatteo_2002}, synchrotron emission from low-redshift galaxy clusters \citep{DiMatteo_2004},etc.  Constructing precise catalogues of the low-frequency radio sky and utilizing them for calibration and foreground removal is a popular method, but still, the latest catalogues cannot entirely suppress the contamination by foregrounds. Other methods include modelling of the foregrounds as a sparse-basis, without the need to actually identify the source of foreground emission in the sky. These methods can lead to over-subtraction, resulting in loss of the signal \citep{ Harker_2009, Tauscher_2018}. Another approach is by sampling the joint posterior between the signal and the foreground, using Bayesian techniques \citep{Sims_2019}. FastICA is a method based on independent component analysis (ICA), which assumes that the foreground components are statistically independent. For a detailed description we direct the readers to \citet{Chapman_2012}. FastICA provides a foreground subtraction method, in which it allows the foregrounds to choose their own shape instead of assuming a specific form a priori. This  type of 'blind source separation (BSS) method' has been also applied successfully to other cosmological observations, such as in CMB foreground removal. GMCA is another method based on independant component analysis, which uses diversity in morphology to separate out the various components. A comparision between the various methods of foreground mitigation can be found in \citet{Chapman_2019}. A recent method was proposed by \citet{mertens_2018}, claiming that foreground modelling via Gaussian processes, might be able to recover the low \kk-modes of the power spectrum measurements of the EoR signal. However, recent power spectrum limits at $z=9.1$ by \citet{Mertens_2020} demonstrate that the Gaussian Process Regression (GPR) based algorithm is also prone to signal loss, and there needs to be improvements in the signal processing chain to reduce the losses. All these above mentioned methods require a priori assumptions about the foreground model to perform the analysis. 

Over the past few years, applications of machine learning (ML) techniques in various aspects of cosmology, astrophysics, statistics and inference and imaging, have come into being. For example, \citet{Schmit_2017}, developed an emulator for the 21-cm power spectrum using artificial neural networks (ANN) for an wide range of the parameters. This emulator has been further used to constrain the EoR parameters using the 21-cm power spectrum within a Bayesian inference framework. \citet{tiwari2021} have extended the same idea and developed an ANN based emulator for 21-cm bispectrum, a higher order statistics to quantify the high and time evolving non-Gaussianity in the signal \citep{majumdar2018,majumdar2020,kamran2021a,kamran2021b}. They have further used this bispectrum emulator within a Bayesian framework to demonstrate that bispectrum can put better constraints on the EoR parameters compared to power spectrum. 
\citet{Cohen_2019} developed an emulator for the 21-cm Global signal using ANN, that connects the astrophysical parameters to the predicted global signal. Convolutional Neural Networks (CNN) was implemented by \citet{Hassan_2019} to identify reionization sources from 21-cm maps. Deep learning models have been used to emulate the entire time evolving 21-cm brightness temperature maps from the epoch of reionization in \citet{Chardin_2019}, where the authors have further tested their predicted 21-cm maps against the brightness temperature maps produced by the radiative transfer simulations. \citet{Gillet_2019} have recovered the  astrophysical parameters directly from 21-cm images, using deep learning with CNN. \citet{Jennings_2019} have compared machine learning techniques for predicting 21-cm power spectrum from reionization simulations. \cite{Li_2019} have implemented a convolutional de-noising autoencoder (CDAE) to recover the EoR signal by training on SKA images simulated with realistic beam effects. While the above examples are focussed on the  imaging domain, \citet{Choudhury_2020} have used ANN to predict the astrophysical information by extracting the parameters from 21-cm Global signal in the presence of foregrounds and instrument response. \citet{Choudhury_2021} have incorporated  physically motivated 21-cm signal and foreground models, and have used ANN to extract the astrophysical parameters from 21-cm Global signal observations directly. They have also applied their ANN to extract the 21-cm parameters from the EDGES data.   
In an earlier work, \citet{Shimabukuro_2017} used ML algorithms for extracting the parameters of the 21-cm power spectrum, however they have not considered foregrounds in their analysis.  

The extraction of astrophysical and cosmological information is heavily dependent on accurate modelling, parametrization, and use of very well-calibrated instruments. In this paper, we propose an alternate method of signal extraction using ANN. We demonstrate that a very basic ANN based framework can be formulated to extract the \hi~ 21-cm power spectrum from the total observed sky power spectrum directly. 

The structure of this paper is as follows. We start with a basic overview of the 21-cm brightness temperature fluctuations and expressions for the power spectrum
of the EoR redshifted 21-cm signal, and describe the details of the semi-numerical simulations that we have used to generate the realizations of the EoR 21-cm signal, in Section \textsection~\ref{simulationPS}. In Section \textsection~\ref{obsChallenges}, we describe the challenges faced while trying to detect this faint signal and explain how the foreground power spectrum is generated. We also briefly describe how the noise power spectrum is computed. Following this, we describe our  artificial neural network based framework, and elaborate on how we construct the training and test sets in Section \textsection~\ref{ann}. In Section \textsection~\ref{results}, we present our results: the predicted signal parameters from 21-cm power spectrum simulations and the recovered power spectrum from the simulated mock observations. Finally, in Section \textsection~\ref{discussions}, we discuss and summarize our results and elaborate on the scope for future work.
\begin{figure}
    \includegraphics[width=3in]{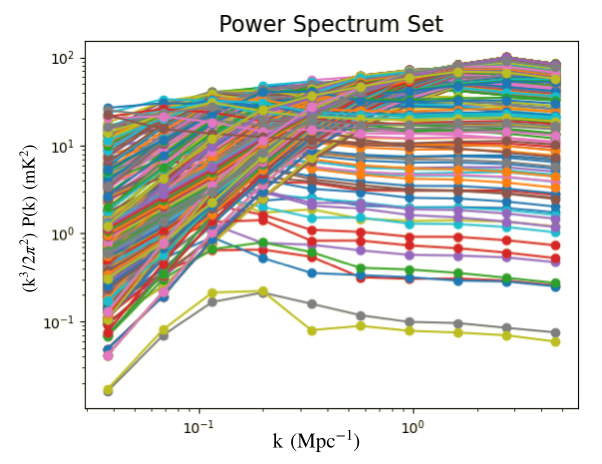}
    \caption{The set of 21-cm power spectra at z=9.01, which are generated by varying the reionization parameters: (a) the ionizing efficiency, $\zeta$,; (b)  the minimum halo mass, $\rm{M_{h.min}}$, within our chosen range. }
    \label{fig:ps set}
\end{figure}
 \section{The 21-cm signal}
\label{simulationPS}
\label{reionyuga}
The hyper-fine splitting of the lowest
energy level of hydrogen, gives rise to the rest-frame $\nu_{21} = 1.42$ GHz radio
signal corresponding to a wavelength of 21 cm. The 21-cm signal is produced by the neutral hydrogen atoms in the IGM, collectively observed as a contrast against the background CMB radiation (\citet{Pritchard_2012} gives a detailed review).
This contrast is the differential brightness temperature, which is the primary observable in 21-cm experiments, given by: 
\begin{equation}
\begin{split}
\delta T_{b}(\nu) & =\frac{T_{s}-T_{\gamma}}{1+z}(1-\exp^{-\tau_{\nu_{0}}})\\
& \approx 27x_{HI}(1+\delta_{b})\left(\frac{\Omega_{b}h^{2}}{0.023}\right)\left(\frac{0.15}{\Omega_{m}h^{2}} \frac{1+z}{10}\right)^{1/2}\\
& ~~\left(1-\frac{T_{\gamma}(z)}{T_{s}}\right)\Big[\frac{\partial_{r} v_{r}}{(1+z)H(z)}\Big]^{-1} ,
\end{split}
\label{eq:global}
\end{equation}
where $x_{HI}$ is the neutral fraction of hydrogen, $\delta_{b}$ is the fractional over-density of baryons, $\Omega_{b}$ and $\Omega_{M}$ are the baryon and total matter density respectively, in units of the critical density, $H(z)$ is the Hubble parameter and $T_{\gamma}(z)$ is the CMB temperature at redshift z, $T_{s}$ is the spin temperature of neutral hydrogen, and $\partial_{r} v_{r}$ is the velocity gradient along the line of sight.
The underlying cosmology and rich astrophysics associated with the evolution of the 21cm signal, makes it a very promising probe into the less explored phases of the evolution of the Universe. It is a prospective tool which will enable us to characterize the formation and the evolution of the first astrophysical sources and, potentially, properties of dark matter across cosmic time. Most recent experiments target to measure the 21-cm signal  either as a sky averaged Global signal or as a power spectrum.
The 21-cm power spectrum from the epoch of reionization can be quantified using the fluctuations of the brightness temperature. 
\begin{equation}
    \rm{\delta T_b(\nu) = T_b(\nu) - \bar T_b(\nu)}.
\end{equation}
The spatial fluctuations of the 21-cm signal in a volume can be decomposed into Fourier modes, given by:
\begin{equation}
    \rm{\delta T_b(\bar \nu)} = \int \frac{d^3k}{(2\pi)^3}e^{(i \bar\kk\cdot \bar r)} \rm{\delta T_b (\bar k)}.
\end{equation}
The expectation value of this quantity $\mathrm{\delta T_b (k)}$ can be expressed as:
\begin{equation}
    \rm{<\delta T_b(\bar \kk)~\delta T_b^*(\bar \kk')>}=\rm{ (2\pi)^3\delta(\kk-\kk')P_{21}(\bar \kk)}
\end{equation}
Here, $\rm{P_{21}(\kk)}$ is the 21-cm power spectrum, which represents the fluctuations of the brightness temperature and tells us about its statistical properties.  The dimensionless power spectrum of the brightness temperature, is given by $\rm{\Delta^2_k=\kk^3P(k)/2\pi^2}$. Throughout this paper, we work with the dimensionless power spectrum, $\rm{\Delta^2}$, also interchangeably referred to as the `Signal PS' in units of $\rm{mK^2}$ (see Fig.~\ref{fig:ps set}).

\subsection{Simulating the 21cm signal from the EoR}
Modelling the 21-cm signal involves taking into consideration various details of the reionization process. These include answering questions such as: when and how did reionization start; how long did it last; was it uniform or patchy; what were the main sources of ionizing photons; etc.  While analytical models can
be very useful to quickly generate the 21-cm signal with a high
dynamic range, they have a disadvantage of not being able to properly deal with the spatial distribution of the reionization process. Numerical models, are able to provide a better improved description of the 21-cm signal, also taking care of the spatial distribution of the associated fields, the disadvantage being that they are comparatively quite slow to run.  Hybrid models are based on a combination of the best of both these methods, and are faster and also able to deal with the spatial resolution of the fields to a large extent.  There are a number of publicly available semi-numerical codes to simulate the evolution of the 21-cm signal, such as 21cmFAST \citep{Mesinger_2011} and SimFAST21 \citep{Santos_2010}. In the publicly available work by \citet{Mesinger_2016}, the Evolution of 21cm Structure (EoS), full simulations (spanning across CD-EoR) considering spin temperature fluctuations have been carried out using two different galaxy models. Considering a maximally optimistic scenario, where the foregrounds have been cleaned very efficiently, they have predicted the duration of reionization. The semi-numerical code used in this paper (ReionYuga, described in the following paragraph) focuses specifically on the EoR.

In this work, we have used a semi-numerical code  \texttt{ReionYuga}{\footnote{\href{https://github.com/rajeshmondal18/ReionYuga}{https://github.com/rajeshmondal18/ReionYuga}}} \citep{Majumdar_2014, Majumdar_2015, Mondal_2015} to generate coeval ionization cubes at redshift $z=9.01$, considering an inside-out model of reionization (which closely follows \citet{Choudhury_2009}). This model is primarily based on an assumption that $\hi$ follows the underlying matter density contrast. The dark matter haloes, which have a masses above a certain threshold ( $\rm{M_{h,min}}$), are the hosts of the sources producing ionizing radiation. To begin with, a dark matter (DM) distribution is generated using a particle-mesh N-body code. The corresponding N-body simulation has $4288^3$ grids with a grid-spacing of $\rm{0.07~Mpc}$ and uses DM particles each of mass $\rm{1.09\times 10^8 M_{\odot}}$. In the following step, a Friends-of-Friends algorithm is used to identify halos in those dark matter distributions, and a halo catalogue is prepared. In the final module, an ionization field is produced using an excursion set formalism \citep{Furlanetto_2004A}, which closely follows the assumption of homogeneous recombination \citep{Choudhury_2009}. The $\hi$ distribution is mapped by those particles whose neutral Hydrogen masses are calculated from the neutral Hydrogen fraction $\rm{x_{\hi}}$, interpolated from its eight adjacent grid points. Following this, for each coeval cube, the positions, peculiar velocities and the H1 masses of these particles are saved.  There are three ionization parameters which we can vary in this prescription, which are:
\begin{itemize}
    \item $\rm{M_{h,min}}$ : Minimum halo mass, which represents the lower limit of the mass of a halo which can collapse and cool down to form the first generation of sources. This cut-off is decided by various cooling mechanisms which are taken into consideration in the model. The value of this parameter decides the onset and duration of the reionization process. 

    \item $\rm\zeta$, Ionizing efficiency: This parameter takes into account a combination of the mostly degenerate astrophysical parameters like the escape fraction of ionizing photons $\rm{f_{esc}}$,the star formation efficiency $\rm{f_{*}}$ and the recombination rate $\rm{N_{\rm rec}}$ \citep{Choudhury_2009} in the prescription for estimating the number of ionizing photons $\rm{N_\gamma}$. For a halo with mass $\rm{M_h\geq M_{h,min}}$, $\rm N_\gamma$ can be written as:
    \begin{equation}
            \rm{N_\gamma(M_h \geq M_{h,min})=\zeta\frac{\Omega_{b}}{\Omega_{b}}\frac{M_h}{m_p}},
    \end{equation}
    where, $\mathrm{\zeta \sim f_* \cdot f_{esc}  * n_{\gamma}* m_p}$, $\Omega_{m}$, $\Omega_{b}$ and $\rm m_p$ are the dark matter density parameter, baryon density parameter and mass of proton respectively. A larger value of $\rm \zeta$ should imply that more efficient ionizing sources are available and hence a larger $\rm N_\gamma$, which would accelerate the process of reionization.
   
    \item $\rm{R_{mfp}}$, mean free path of the ionizing radiation: The ionizing photons travel a certain distance from the point where they were produced, ionising the intervening medium. Thus $\rm{R_{mfp}}$ determines the size of the ionized regions. Varying this parameter, keeping other parameters fixed, has minimal effect on the shape of the power spectrum \citep{Greig_2015, Park_2019}. So we do not include this parameter in our list of target output parameters.
\end{itemize}
\subsection{Preparing the data by varying the ionization parameters}
Our simulations were performed on a cubic box of comoving volume $\rm 215~Mpc^{3}]$, at redshift, $z=9.01$, using \texttt{ReionYuga}. To generate the gallery of 21cm power spectra required for our work (see Fig.~\ref{fig:ps set}), we have varied the ionizing efficiency, $\zeta$ in the range $[18-200]$, and the minimum halo mass, $\rm{M_{h,min}}$ in the range $[10.87 \times 10^9 - 1.195 \times 10^{11} \rm M_{\odot}]$, generating different reionization histories ( see Tab.\ref{Tab:1} for the range of parameters).Each of the 21cm power spectrum in Fig.~\ref{fig:ps set} corresponds to a different combination of the parameters $(\zeta,\rm{M_{h,min}})$, gridded uniformly across the mentioned range.
We choose our redshift of interest around the estimated midpoint of reionization, where the 21-cm signal is expected to peak and the Universe is assumed to be approximately $50\%$ ionized. For our work, we choose the $k$ modes to be in the range, $\rm{0.17 <k <0.35~Mpc^{-1}}$. Most current and upcoming EoR experiments aim to observe the 21-cm signal signal in this redshift regime, with better sensitivities at lower \kk-modes. We use this set of 21-cm power spectra to construct the various training and test sets for ths work, which is explained in detail in \textsection~\ref{results}.

\begin{table}
\centering
\begin{tabular}{|c|c|}
     \hline
     Parameter & Range \\
     \hline
     $\rm{\zeta}$ & $\rm{18-200}$\\
     $\rm{M_{h,min}}$ & $\rm{1.09 \times 10^{9}-1.19 \times 10^{11} M_{\odot}}$\\
     \hline
\end{tabular}
\caption{Range of values of the parameters used to generate different ionization histories}
\label{Tab:1}
\end{table}

\section{Observational Challenges}
\label{obsChallenges}
One of the major challenges in the detection of the 21-cm signal are the bright galactic and extragalactic foregrounds. These foregrounds are primarily constituted of synchrotron radiation, which are several orders of magnitude brighter than the \hi~ signal. It is extremely challenging to model and
subtract out the sources constituting the foregrounds, due to ionospheric fluctuations, lack of detailed knowledge of the instrument and systematics.
Strategically, there could be two ways of dealing with this issue. One is to model the foregrounds very well and subtract it from the observed sky signal, while another would be to opt for foreground avoidance. A lot of effort has been made in the past decade on foreground
removal for detecting the 21-cm power spectrum from EoR. For example, \citet{Morales_2005a, Jelic_2008, Liu_2009a, Liu_2009b, Chapman_2012, Paciga_2013} have discussed in detail, the foreground subtraction technique. In this approach, a foreground
model (empirically obtained form prior observations) is subtracted from the data and the residual data is used to compute the 21-cm
power spectrum. In contrast, foreground avoidance \citep{Datta_2010a, Parsons_2012, Pober_2013, Ali_2015, Trott_2016} is an alternative approach based on the idea that contamination from any foreground
with smooth spectral behaviour is confined only to a wedge in cylindrical $(k_\perp, k_\parallel)$ space due to chromatic coupling of an interferometer with the foregrounds. The \hi~ power spectrum can be
estimated from the uncontaminated modes outside the wedge region termed as the EoR window
where the \hi~ signal is dominant over the foregrounds.  With their merits and demerits, these two
approaches are considered complementary \citep{Chapman_2016}.

Developing algorithms and techniques to characterize and remove the foregrounds efficiently, are the major goals of most of the current and future radio-telescopes. Generally, foreground sources have a smooth power-law continuum spectra. Averaging over many such sources with different spectral indices and spectral structure would yield a smooth power-law foreground. Owing to this fact, detection of the spectral structure corresponding to a distribution of regions containing neutral and ionized hydrogen, would be feasible. The foreground power spectrum is usually represented as an angular power spectrum in $\ell$. Since we have considered coeval simulation boxes at redshift 9.01 only, and are interested in the spherically averaged power spectrum, we need to represent the foreground power spectrum also in terms of \kk. In the following subsection, we will describe how we have simulated the foreground power spectrum for our work.   


\subsection{Simulating the foreground PS}
\label{Foregrounds-sim}
The foreground power spectrum is usually modelled as a power law, in both $\ell$ and $\nu$. It can be expressed as \citep{Santos_2005}:
\begin{equation}
    \mathrm{C_\ell\approx A (\ell/1000)^{-\beta} (\nu_f/\nu)^{2\alpha}}.
\end{equation}

The spectral index $\alpha$ depends on the energy distribution of relativistic electrons, and it varies slightly in the sky, with a spectral steepening at larger frequencies. Following \citet{tegmark1997}, the mean value of $\alpha$ is  usually taken to be $2.8$. The $408 ~\rm MHz$ Haslam map suggests that the power law index, $\beta$, varies approximately between $2.5-2.8$. From \citep{Santos_2005}, we take the value of amplitude A to be $700$. We have varied $\alpha$ in the range $2.75-2.85$ and $\beta$ in the range $2.20-2.79$ to generate the set of foreground power spectra for this work. Our choice of the range of $\beta$ is motivated from actual observations of the ELIAS N1 field \citep{Chakraborty_2019}.
The power spectrum of the fluctuations at two different frequencies can be written as: 
\begin{equation}
    C_\ell(\nu_1,\nu_2) \equiv \langle a_{\ell m}(\nu_1)a_{\ell m}^*(\nu_2) \rangle.
\end{equation}
In the flat-sky approximation, following the formulation in \citet{Mondal_2018, Datta_2007},  P(k) is the Fourier transform of $C_\ell (\Delta \nu)$, and can be expressed as,
\begin{equation}
    \rm{P(k_\perp, k_\parallel) = r_\nu^2 r_\nu' \int d(\Delta \nu) e^{-ik_{\parallel}r_{\nu}'\Delta\nu} C_\ell(\Delta\nu)},
\end{equation}
where, $\rm{k_\parallel}$ and $\rm{k_\perp= \ell/r_{\nu}}$ are the components of  $\rm{\kk=\sqrt{k_{\parallel}^2+(\ell/r_{\nu})^2}}$, which are parallel and perpendicular to the line of sight, respectively; $r_{\nu}$ is the comoving distance given by: $\rm{r_{\nu}=\int_0^z dz'[c/H(z')]}$. Following \citet{Datta_2007,Mondal_2018}, and solving this integration for P(\kk), we convert the foreground angular power spectrum to $\rm{P_{FG}(\kk)}$.
\subsection{Simulating the Noise PS}
In this subsection, we briefly describe the formalism to obtain the sensitivity to the 21-cm power spectrum, and direct the readers to \citet{Pober_2013, Parsons_2012}, for details. In this method, a $\textit{uv}$ coverage of the observation is generated by gridding each baseline into the $\textit{uv}$ plane and including the effects of the Earth's rotation over the course of the observation. The sensitivity to any mode of the dimensionless power spectrum is given by: 

\begin{equation}\label{eq16}
     \rm{\Delta^2_N(k) \approx X^2Y \frac{k^3}{2\pi^2} \frac{\Omega}{2t} T^2_{sys}},
\end{equation}
where t is the integration time for sampling a particular $(u,v,\eta)$-mode, and the factor of two in the denominator comes from the explicit inclusion of two orthogonal polarization components to measure the total unpolarized signal. $\Omega$ is a factor related to the solid angle of the primary beam and $T_{sys}$ is the system temperature. $X^2Y$ is a cosmological scalar converting $(u,v)$ units into comoving wavenumber units.  In this work, we have used a slightly modified version of \textsc{21cmSense} \citep{Parsons_2012,Pober_2013} to obtain the $\textit{uv}$-coverage for instruments HERA and SKA (actually \textsc{SKA1 LOW}), by using the corresponding calibration files. We have obtained the thermal noise power spectra for these experiments corresponding to a total of 1080 hours of observation for SKA (in 6 hours of tracking mode observation per day, for 180 days) and HERA (1080 hours in drift mode), for this work.  

Now that we have the simulated signal, foreground and the noise power spectrum ready, we now describe our ANN framework and the construction of the corresponding training and test sets in the following sections.

\section{A Neural Network Framework to extract the 21-cm PS}
\label{ann}
Artificial neural networks (ANN) are one of the several available machine learning techniques which can be used for developing supervised learning-based frameworks. ANN are capable of mapping associations with the given input data and the target parameters of interest, without the knowledge of their explicit analytical relationships.
A basic neural network has three kinds of layers: an input layer, one or more hidden layers, and an output layer. Each neuron in a layer is connected with every neuron in the next layer, and the connection is associated by a weight and a bias. During the training process, the network optimises a chosen cost function by repeatedly back-propagating the errors and adjusting the weights and biases. A chunk of the data is put aside as the validation dataset and the performance of the network is validated using this set. 
For a detailed description of the training process, please see \citet{Choudhury_2020}. Once training and validation is completed, a test set is used to predict the output parameters. We have used publicly available python-based packages \textsc{scikit learn} \citep{Pedregosa_2011} and \textsc{keras} in this work to design the network.\\
In this paper, we compute R2-scores for each of the output parameters as the performance metric. The R2-score is defined as:

\begin{equation}
\mathrm{R^2=\frac{\Sigma(y_{pred}-\overline{y}_{ori})^2}{\Sigma(y_{ori}-\overline{y}_{ori})^2}=1-\frac{\Sigma(y_{pred}-y_{ori})^2}{\Sigma(y_{ori}-\overline{y}_{ori})^2}}
\end{equation}
where, $\overline{y}_{\mathrm{ori}}$ is the average of the original parameter, and the summation is over the entire test set. The score $\rm R^2 =1$, implies a perfect inference of the parameters, while $\rm R^2$ can vary between 0 and 1. \\
Training of artificial neural networks (ANN) in a supervised manner is very target-specific, and has very different network architectures for different applications. While we use simple multi-layer perceptron type of ANN in this work and also in the previous ANN-based works\citep{Choudhury_2020,Choudhury_2021}, the methodology and the ANN architecture to extract the parameters from the observations of the sky-averaged 21-cm Global signal is very different from our implementation to the power spectrum synthetic-observations.
The parameter extraction process becomes less straightforward when we have to deal with the extraction of the faint HI 21-cm power spectrum parameters, taking into consideration fluctuations in the foreground.
While ML methods are extremely fast and computationally very efficient, the ANN model performances need to be carefully analysed. There can be issues of "overfitting", which is typically the case when the trained model fails to generalize what it has learned during the training process. Such a scenario can be overcome by suitable regularization techniques. There can also be an "underfitting" scenario, where the training error does not reduce even after several iterations. This can also be solved by making the network more complex (e.g: by adding more layers) and also by using a well represented training datasets. The ANN is trained to find an association or mapping between the input data and the target features. In parameter estimation, the use of ANN allows us to to explore a large parameter space conveniently without requiring a specified prior and also by-passes the calculation of the likelihood, expediting the process tremendously. \\
In the following section, we first describe the basic architecture of the ANN and elaborate on the details of how the training and test datasets were created. We then explain the ANN framework which we have used to extract the \hi~ PS from the synthetic datasets, and discuss the results that we have obtained.

\begin{figure*}
    \centering
    \includegraphics[width=6.75in]{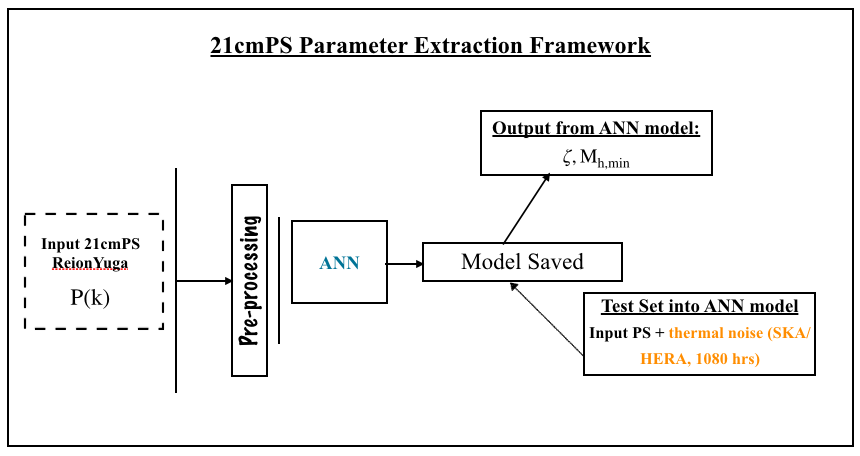}
    \caption{This flowchart describes the framework used to predict the parameters $\rm{\zeta,M_{h,min}}$ from the mock datasets from which the foregrounds are assumed to have been perfectly removed.}
    \label{fig:params-framework}
\end{figure*}
\section{Results}
\label{results}
In this work, we have used a basic multi-layer perceptron neural network to calculate the parameters from the \hi~ power spectrum data. The number of input neurons of the ANN corresponds to the dimension of each of the input power spectra. In our study, we have restricted our analyses to the \kk-modes where we expect the sensitivity of the upcoming telescopes to be optimal. Hence, we have considered only $0.17<\rm \kk <0.35~Mpc^{1}$ for the upcoming experiments like the HERA and the SKA. Our training sets consists of 1148 sample 21-cm PS (see Fig.~\ref{fig:ps set}). Using the simulated signal power spectrum, the foreground, and the thermal noise corresponding to the instrument in question, we construct datasets to train the artificial neural network. 

\begin{table}
    \centering
     \begin{tabular}{|c|c|c|}
     \hline
     Test Sets & $\mathrm{\zeta}$ & $\mathrm{log~M_{h,min}}$ \\
     \hline
     No-noise & 0.946 & 0.993 \\
     \hline
     SKA & 0.846 & 0.868 \\
     \hline
     HERA  & 0.774 & 0.830   \\
     \hline
    \end{tabular}
\caption{Case 1: Training set does not contain added foregrounds or noise, while the test set contains noise corresponding to 1080 hours of observation of HERA and SKA. The calculated R2 scores are listed in this table for HERA and SKA. However, when the test sets do not contain any added noise, the R2 scores are $0.993$ and $0.946$ for $\rm{log(M_{h,min})}$ and $\rm{\zeta}$ respectively.}
\label{tab:tab2}   
\end{table}

\begin{table}
    \centering
     \begin{tabular}{|c|c|c|c|c|c|}
     \hline
       & 21 cm  & Fore- & Noise & Noise & ANN \\
            & signal & -ground & (SKA) & (HERA) &  \\
     \hline
      1a & \checkmark & $\times$ & $\times$ & $\times$ & Single \\
     \hline
     1b & \checkmark & $\times$ & \checkmark & $\times$ & Single  \\
     \hline
     1c & \checkmark & $\times$ & $\times$  & \checkmark & Single  \\
     \hline
     2a & \checkmark & \checkmark & $\times$  & $\times$ & Dual  \\
     \hline
     2b & \checkmark & \checkmark & \checkmark & $\times$ & Dual \\
     \hline
     2c & \checkmark & \checkmark & $\times$ & \checkmark & Dual \\
     \hline
    \end{tabular}
\caption{In this table, we summarise all the various cases that we have explored in this paper.}
\label{tab:tab22}   
\end{table}

\subsection{Case 1: Only 21-cm Signal without foregrounds}
\label{TrainingWithNoFG}
\begin{figure*}
\includegraphics[width=6.25in]{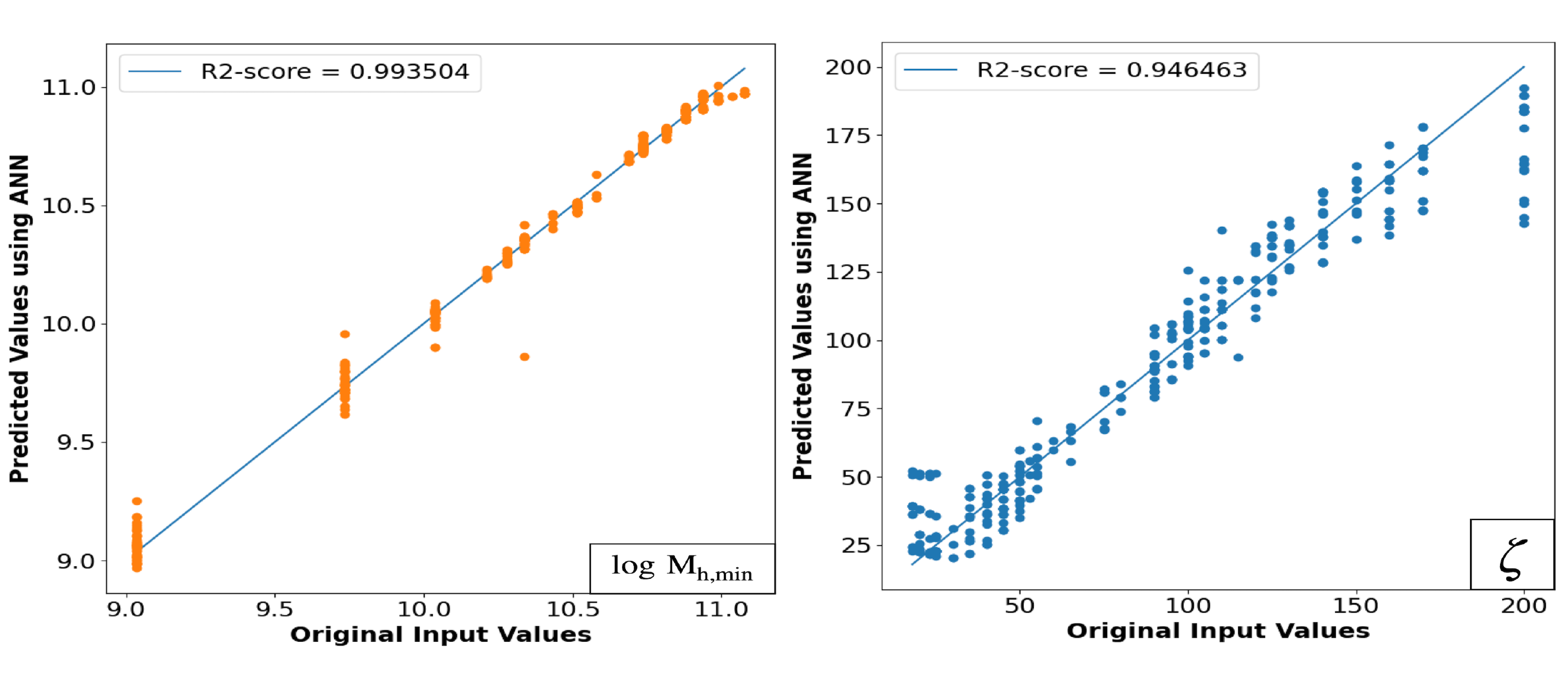} 
\caption{Case 1a: Predicted values of the astrophysical parameters $\rm{\log~M_{h,min}}$in $\rm{(M_{\odot})}$ (left) and $\rm \zeta$ (right), along with the computed R2-scores, from the test set which does not contain any added noise or foregrounds.}
\label{fig:nofg}
\end{figure*}

\begin{figure*}
\includegraphics[width=6.5in]{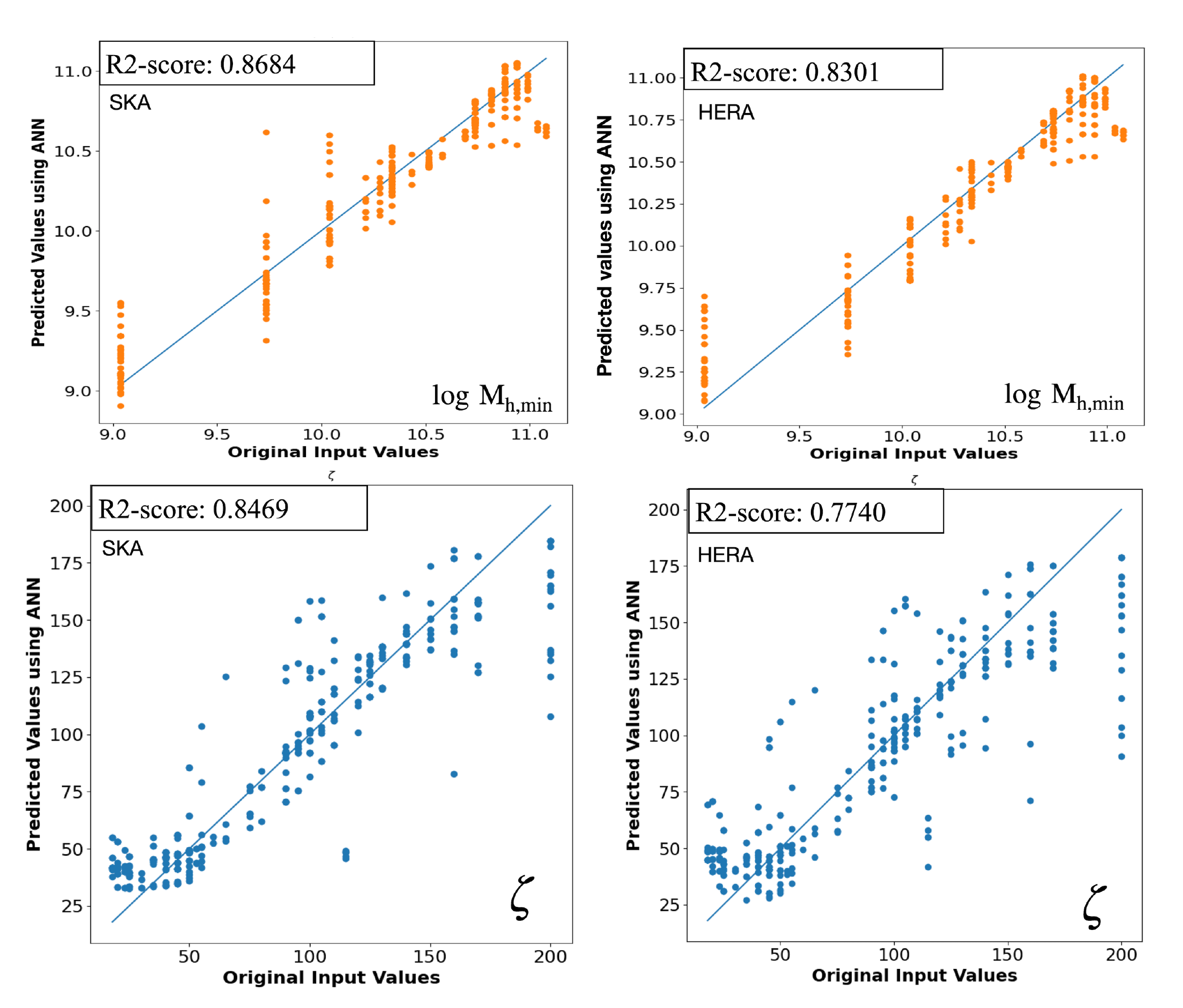}
\caption{The above plot shows the original versus predicted values of the parameters, $\zeta$ and $\rm{M_{h,min}}$ in $\rm M_\odot$, from the ANN when the test sets contains only thermal noise corresponding to 1080 hours of observations of SKA (Case 1b) and HERA (Case 1c) respectively, in addition to the signal. The R2 score calculated for each of the parameters are on the top left hand corner box for each plot. We see that the parameters are recovered much more accurately from the case where SKA-noise has been added, as compared to HERA.}
\label{fig:added-noise-nofg}
\end{figure*}

Firstly, we demonstrate a basic ANN framework to predict the astrophysical parameters from the \hi~ power spectrum (Fig.~\ref{fig:params-framework}). Such an implementation was earlier demonstrated by \citet{Shimabukuro_2017}.  This is a proof-of-concept check to see how well the parameters can be extracted from the \hi~ power spectrum data. Each sample in the training dataset for this case, consists of only the $\hi$ power spectra and is constructed as follows:
\begin{equation}
    \rm{PS_{tot}(k)} = PS_{\hi}(k) ;
\end{equation}
where, $\rm{PS_{tot}(k)}$ is the total power spectrum. The ANN consists of an input layer, 3 hidden layers (activated by tanh, elu, tanh respectively) and an output layer with a linear activation function. We have used 'mean squared error'('mse') as the loss function and 'adagrad' as the  optimizer \citep{Duchi_2011} for this network. Adagrad is a gradient-squared based optimizing algorithm which is used to minimize the loss-function without getting stuck in local minima, saddle points, or plateau regions.
To construct the test datasets, we consider a randomly drawn set consisting of 100 \hi~ power spectra. This \hi~ 21-cm PS set is kept fixed, for all the different test sets constructed in this work. In addition to the \hi~ power spectrum, $\rm PS_{\hi}$, we add thermal noise corresponding to 1080 hours of observation of HERA and SKA, represented by $\rm PS_{noise}$. We construct three different test sets, the total power spectrum ($PS'_{tot}$) being defined as:
\begin{enumerate}
    \item  Case 1a: Signal
    \begin{align*}
    \hspace{10mm} \rm{PS'_{tot}(k)  = PS_{H1}(k)}
    \end{align*}
    \item  Case 1b: Signal + Noise(SKA)
    \begin{align*}
    \hspace{10mm} \rm{PS'_{tot}(k)  = PS_{H1}(k) + PS_{\rm noise, SKA}}
    \end{align*}
    \item Case 1c: Signal + Noise (HERA)
    \begin{align*}
    \hspace{10mm} \rm{PS'_{tot}(k) = PS_{H1}(k)} + PS_{\rm noise, HERA}
    \end{align*}
\end{enumerate}
The noise contribution is computed using a slightly modified version of \textsc{21cmSense}, corresponding to 1080 hours of observation for each of the telescopes.
In Fig.~\ref{fig:nofg}, we have plotted the predicted parameters when the test set does not contain any added noise. In Fig.~\ref{fig:added-noise-nofg}, we have used test sets with the thermal noise corresponding to SKA and HERA, which are shown in each of the panels. We have plotted the minimum halo mass in log-scale, $\rm{M_{h,min}}$. It can be noted from the plots that the R2-scores of the predictions corresponding to the no-noise test sets are much higher than the ones in which the noise corruption has been added, as expected. Table \ref{tab:tab2} summarizes the details of the predictions for each of the cases.

\begin{figure}
    \centering
    \includegraphics[width=\columnwidth]{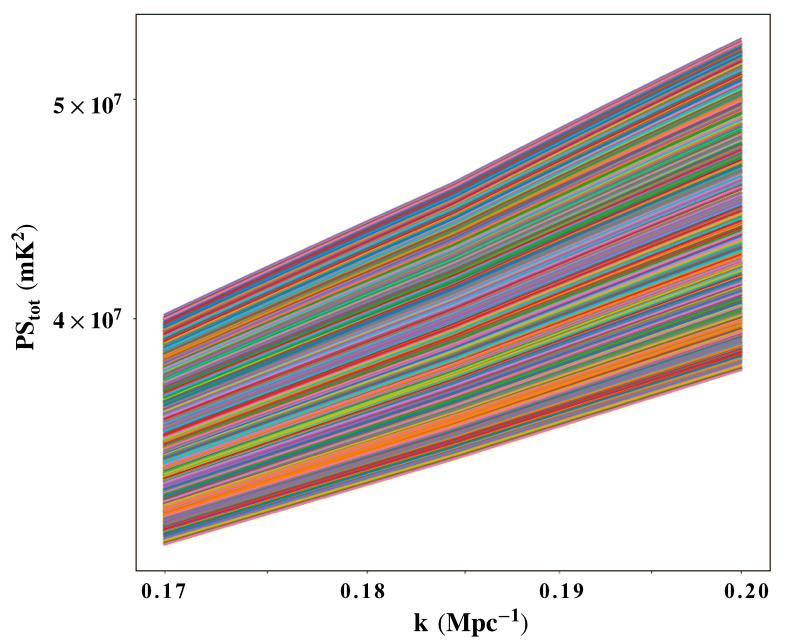}
    \caption{The training dataset for the 2-step ANN framework, constructed by combining the 21-cm Power Spectra and foregrounds. The training set is limited to \kk-modes in the range, $0.17<\kk<0.35\rm{Mpc^{-1}}$, which is suitably chosen to accommodate  the maximum sensitivities corresponding to HERA and SKA.}
    \label{fig:PSwFg}
\end{figure}

\subsection{Case 2: 21-cm signal in presence of foregrounds}
\label{TrainingWithFG}
As we progress to build more realistic training datasets from the upcoming radio-experiments, mock datasets include the effect of the foregrounds as well as the thermal noise corruptions.  
For the training sets, we use the foregrounds that has been generated using the prescription described in \textsection\ref{Foregrounds-sim}. Though it would be more realistic to add the foregrounds in the image domain, and then compute the power spectra to build the training sets, we have assumed that by adding the FG power spectra across the available \kk-modes, we can mimic the total power spectrum from foreground corrupted datasets to some extent. While working with synthetic datasets, the simulated FG maps are ideally added in the image plane, and then the power spectrum is computed. Such a formalism has been followed by \citet{Li_2019} to separate the EoR signal from an EoR+FG field. Though this is a more ideal way of simulating observational effects, we have presented an alternate approach to obtain the power spectra, without going into the imaging domain.  As the flat sky approximation holds good in the scales considered in this work, we assume that adding the FG power spectra per k-mode would also be a good representation of the total power spectrum from foreground corrupted datasets to some extent, without actually going into the imaging domain and introducing more imaging related corruptions. The purpose of making such an assumption is to demonstrate that neural network can be used to recover the \hi~ 21cm power spectra from a foreground dominated dataset without going into the imaging domain. \citet{Mesinger_2016} have used full EoS simulations but have considered only two possible cases: a maximally optimistic scenario, with no effect of foreground or perfect foreground cleaning.\\    
We define the total power spectra as:
\begin{equation}
    \rm{PS_{tot}(k) = PS_{H1}(k) + PS_{FG}(k),}
\end{equation}
here, $\rm{PS_{H1}}$ and $\rm{PS_{FG}}$ are the 21-cm signal power spectrum and the foreground power spectrum respectively. The training datasets for this case are shown in  Fig.~\ref{fig:PSwFg}. We see that the \hi~ signal PS is totally buried in the foregrounds, which are orders of magnitude higher. 
The test sets are constructed in a similar manner, by adding the noise PS corresponding to HERA and SKA as:  
\begin{equation}
    \rm{PS_{tot}(k) = PS_{H1}(k) + PS_{FG}(k) + PS_{noise}(k).}
\end{equation}

When we attempted to train the ANN to predict the associated 21-cm signal parameters directly from these training sets, we were not able to recover the parameters and achieve good accuracy. The best accuracies that we could achieve, were only $\sim 30-35 \%$ for $\zeta$ and $\rm{M_{h,min}}$, that too, for the case where no noise was added to the test sets. As an alternate approach, we have proposed a two step ANN framework to predict the 21-cm signal from the total observed power spectrum. 
\begin{figure}
    \centering
    \includegraphics[width=\columnwidth]{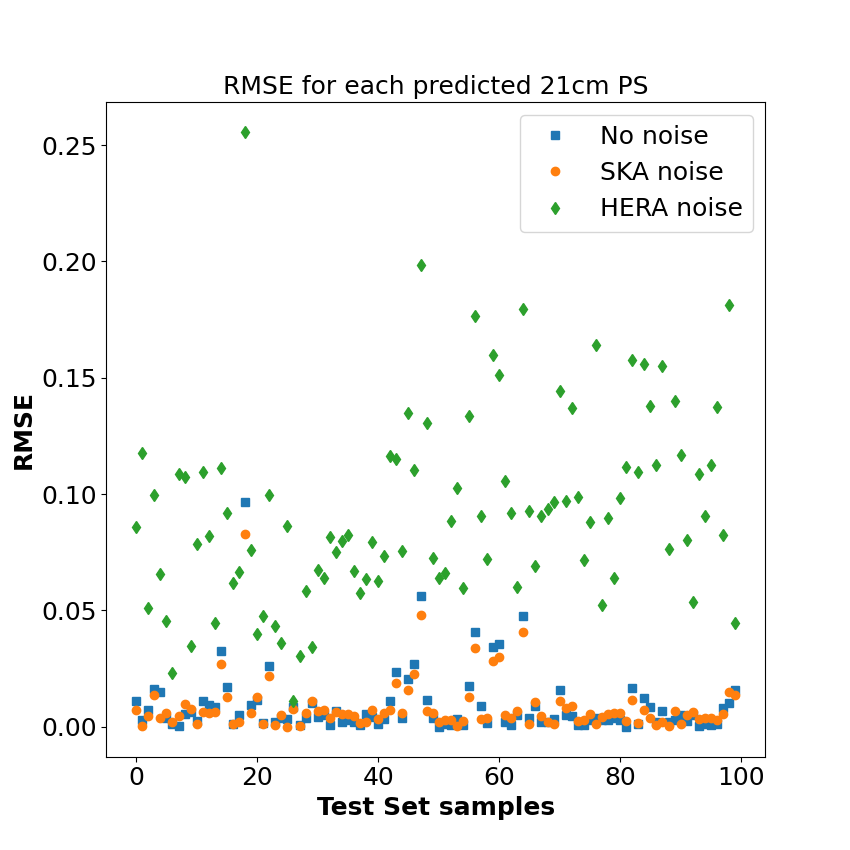}
    \caption{The RMSE for each of the predicted 21-cm PS in the foreground-corrupted test datasets, is calculated and plotted in the above plot. We see that the RMSE corresponding to the case where the test set samples contain noise corresponding to HERA, for 1080 hours of observation is much higher than the cases where the added noise corresponds to SKA (1080 hours of observation) and when no noise is added. 
    }
    \label{fig:rmse}
\end{figure}
\\ \\

\begin{figure*}
    \centering
    \includegraphics[width=6.75in]{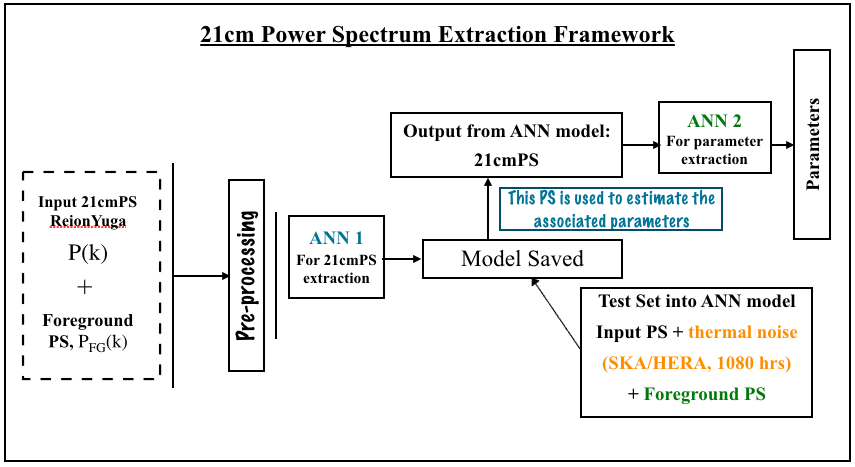}
    \caption{This flowchart describes the two-step ANN framework which can extract the 21-cm power spectrum and the associated reionization parameters from foreground dominated datasets. The framework incorporates two neural networks: ANN1 predicts the 21-cm power spectrum from the total power spectrum.  ANN2 predicts the associated astrophysical parameters from the output of ANN1.}
    \label{fig:ANN-framework}
\end{figure*}

\begin{figure*}
    \centering
    \includegraphics[width=6.75in]{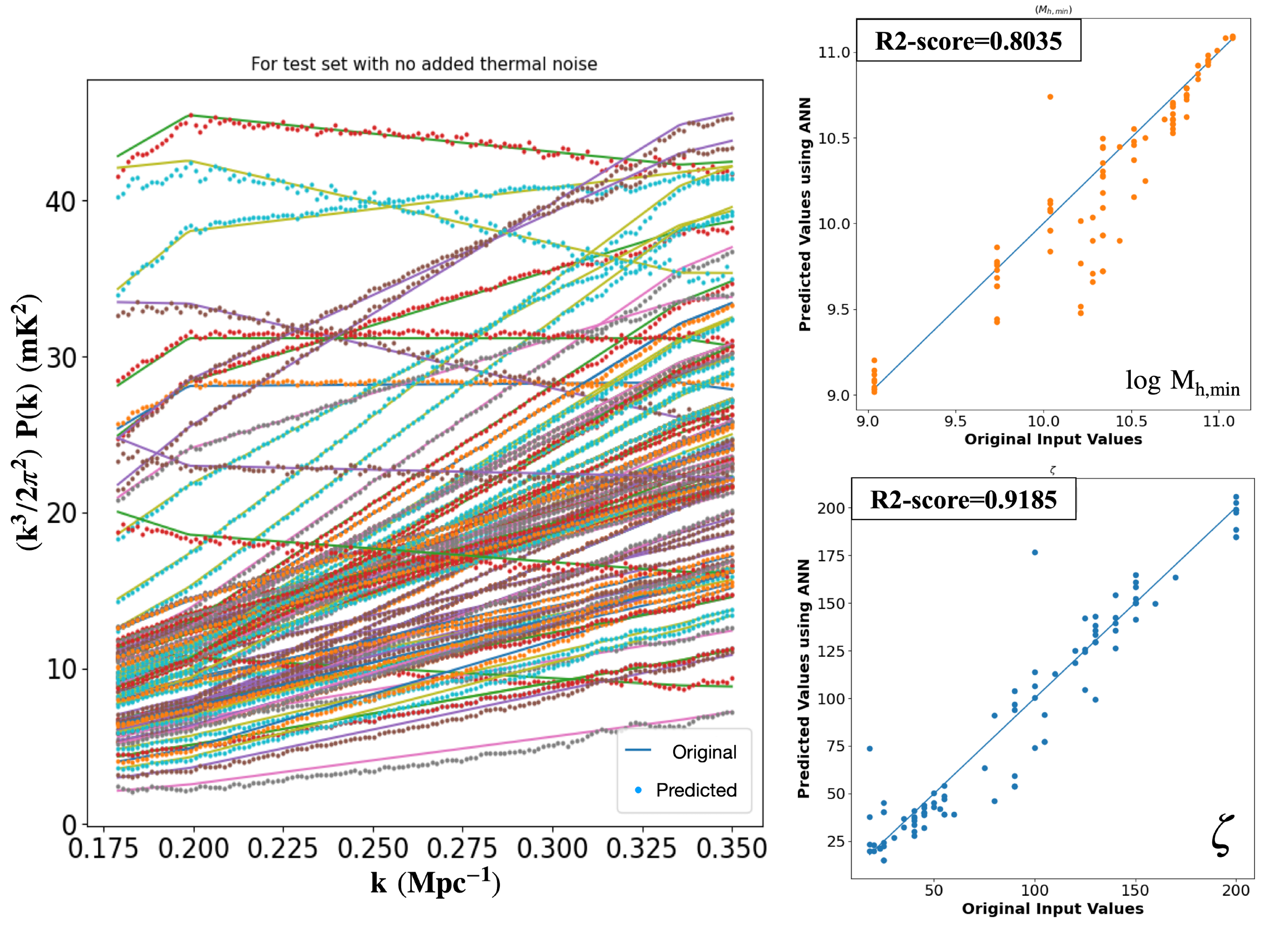}
    \caption{Case 2a: The predictions of the 21-cm power spectra from the foreground-corrupted test set without thermal noise, is shown in the above figure. The left panel shows the output of \texttt{ANN1}, i.e., the original and predicted \hi~ power spectra.  The right panel shows the predicted parameters $\rm\zeta$ and $\rm M_{h,min}$ in $\rm M_\odot$, from \texttt{ANN2}.
    }
    \label{fig:recon_ps_wfg_no_noise}
\end{figure*}

\begin{figure*}
    \centering
    \includegraphics[width=6.75in]{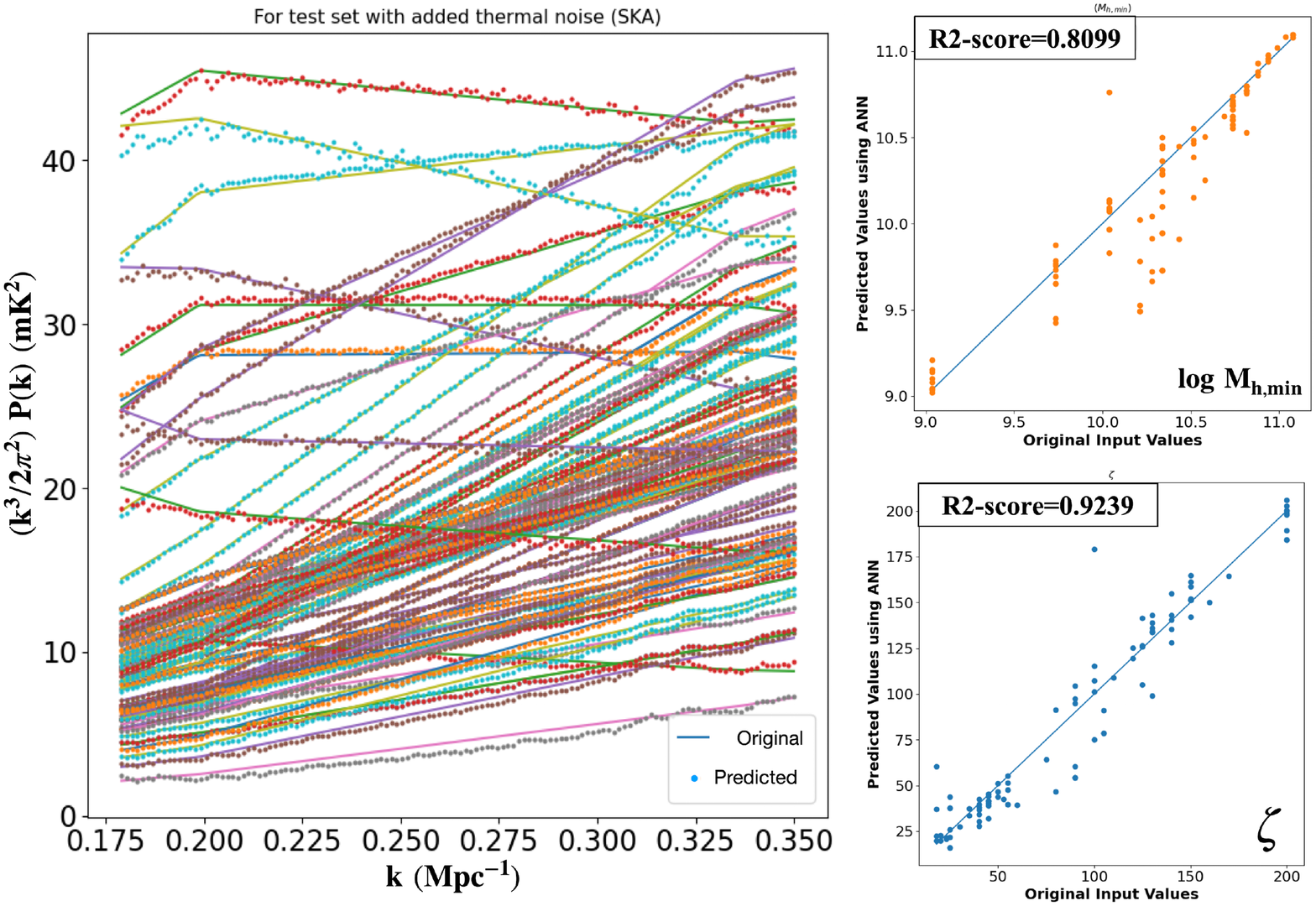}
    \caption{Case 2b: The left panel shows the original and the predicted 21-cm power spectra, from the foreground-corrupted test set, containing thermal noise corresponding to 1080 hours of observations using SKA. The right panel shows the predicted parameters $\rm\zeta$ and $\rm M_{h,min}$ in $\rm M_\odot$, from \texttt{ANN2}
    }
    \label{fig:recon_ps_wfg_ska}
\end{figure*}

\begin{figure*}
    \centering
    \includegraphics[width=6.75in]{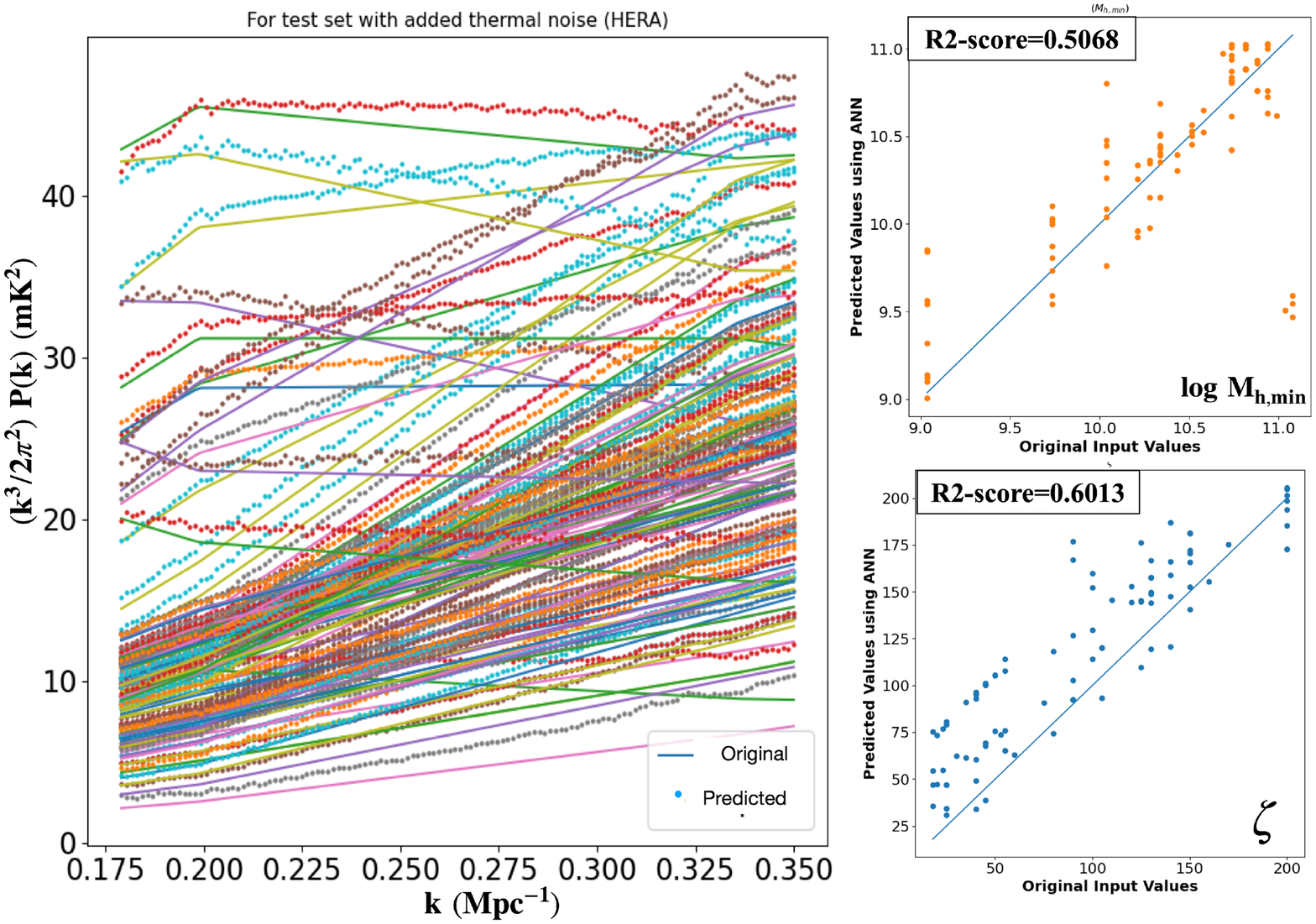}
    \caption{Case 2c: In the left panel, the predicted 21-cm power spectra from the test set corrupted by foregrounds and thermal noise corresponding to 1080 hours of observations using HERA, are shown.
    It can be observed that there is a clear deviation from the original PS, in each of the predicted power spectra from \texttt{ANN1}. The right panel shows the predicted parameters $\rm\zeta$ and $\rm M_{h,min}$ in $\rm M_\odot$, from \texttt{ANN2}.
    }
    \label{fig:recon_ps_wfg_hera}
\end{figure*}

\hspace{10mm}{\textbf  { \textit {Invoking the 2-step ANN framework:}}}  When we attempted to extract the reionizaton parameters directly from the 21cm signal+foreground datasets, the accuracies of the predictions fall considerably. Thus we suggest to perform the training in two segments: in the first step we extract the \hi~ power spectrum from the foreground corrupted datasets. Next, using this as the input into another ANN, we predict the astrophysical parameters (Fig.~\ref{fig:ANN-framework}.
In this framework, at first, we use \texttt{ANN1} to extract the \hi~ 21-cm PS directly from the total sky power spectra. \texttt{ANN1} is a multilayer perceptron, with 128 input nodes, corresponding to the number of \kk-modes we have considered in our training set data. In our model, we use 3 hidden layers, with `tanh' activation functions, with a learning rate of 0.01. The final output layer also contains of 128 nodes. The overall network has a node structure of $128-\{1024-512-216\}-128$. The output from this \texttt{ANN1} is the \hi~ signal power spectrum, that we are interested in. 
We construct three different test sets to check the performance of the ANN-framework:
\begin{enumerate}
    \item  Case 2a: Signal + Foreground
    \begin{align*}
    \hspace{10mm} \rm{PS'_{tot}(k)  = PS_{H1}(k) + PS_{FG}(k)}
    \end{align*}
    \item  Case 2b: Signal + Foreground + Noise (SKA)
    \begin{align*}
    \hspace{10mm} \rm{PS'_{tot}(k)  = PS_{H1}(k) + PS_{FG}(k) + PS _{\rm noise, SKA}}
    \end{align*}
    \item  Case 2c: Signal + Foreground + Noise (HERA)
    \begin{align*}
    \hspace{10mm} \rm{PS'_{tot}(k) = PS_{H1}(k) + PS_{FG}(k) + PS _{\rm noise,HERA}}.
    \end{align*}
\end{enumerate}
Here, $PS'_{tot}$, $PS_{\hi}$ are the total power spectrum and the \hi~ power spectrum respectively. $PS_{noise}$ denotes the thermal noise power spectrum corresponding to SKA and HERA respectively. \texttt{ANN1} predicts the 21-cm PS from the test sets. We have plotted the predictions from \texttt{ANN1} in the left panels of
Figs.\ref{fig:recon_ps_wfg_hera},\ref{fig:recon_ps_wfg_no_noise},\ref{fig:recon_ps_wfg_ska}). 
\\
\hspace{10mm}\textbf{\textit{Computing the performance metrics :}} In order to compare the accuracy of the 21-cm PS predicted by the \texttt{ANN1} for the different test cases, we have computed the RMSE's for each of the samples in our test dataset. We use the relation: $\mathrm{\sqrt{\sum_{i=1}^{N}(y_{o}-y_{p})/y_o)^2/N}}$, to calculate the RMSE. Here, $y_o,y_p$ denotes the original and predicted values of the \hi~ PS at a particular value of \kk. The summation is over the \kk's, so N is the dimension of the power spectrum (which is 128 in our case). This normalised RMSE gives us a measure of the offset or error in the predicted \hi~PS. We also observed that sampling our chosen parameter space (corresponding to the reionization parameters) with latin hypercube sampling (LHS) method did not significantly improve the performance of the ANNs, particularly for the case where training is done with the foreground-corrupted power spectra. The calculated RMSE for each of the samples in the test sets corresponding to the no-noise, HERA-noise and SKA-noise case, are plotted in Fig.~\ref{fig:rmse}. We achieve high accuracies of $\approx~95-99\%$ 

\begin{figure}
    \centering
    \includegraphics[width=3.2in, height=2.75in]{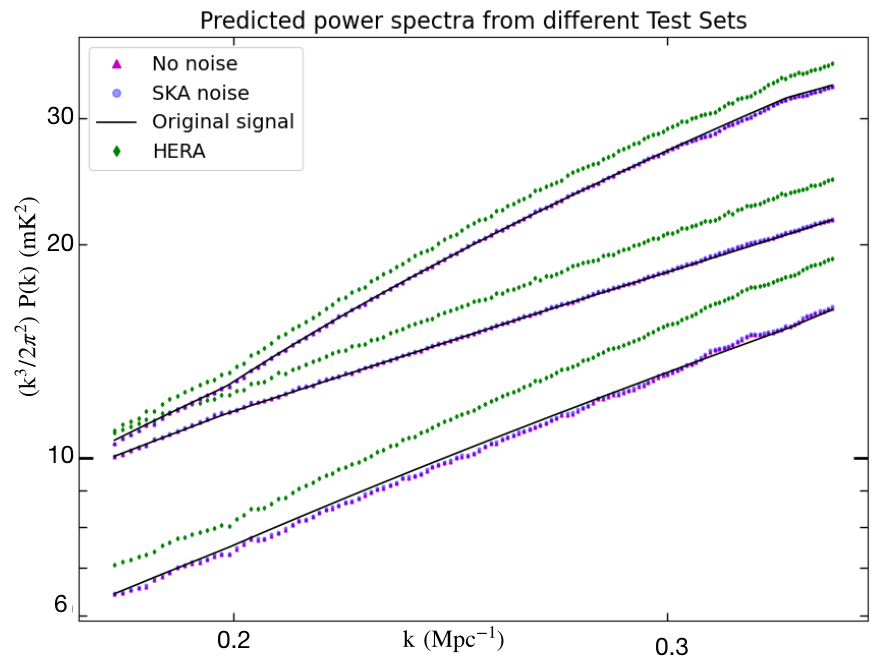}
    \caption{The RMSE for each of the predicted 21-cm power spectra from the foreground-corrupted test datasets is calculated and plotted in the above plot. We see that the RMSE corresponding to the test set samples containing noise corresponding to HERA for 1080 hours of observation is much higher than the test set containing noise corresponding to SKA for 1080 hours of observation. The green and when no noise is added. The x-axis is just a numeric label corresponding to each of the test set samples. }
    \label{fig:ps-recon}
\end{figure}

The predicted 21-cm-Signal power spectra, for all the samples of the test sets is shown in the left panels of Figs.~\ref{fig:recon_ps_wfg_no_noise}, ~\ref{fig:recon_ps_wfg_ska},~\ref{fig:recon_ps_wfg_hera}. The dotted plots represent the predicted power spectra and the solid lines are the original input power spectra. We obtain very good accuracy from the first stage of the framework, \texttt{ANN1}, whose output is then passed into another ANN, which we have labelled as \texttt{ANN2}. \texttt{ANN2} is the same model as discussed in \textsection~\ref{TrainingWithNoFG}, which has been trained for the no-foreground case to predict the parameters associated with the input signal. The flowchart describing the two-step framework is shown in Fig.~\ref{fig:ANN-framework}.The predicted parameters are are shown in the right panels of Figs.~\ref{fig:recon_ps_wfg_no_noise},~\ref{fig:recon_ps_wfg_ska},~\ref{fig:recon_ps_wfg_hera}. We can see that the error in the first step of the framework, is carried over to the prediction of the parameters. This is particularly evident from the test sets containing HERA-noise. The recovered 21-cm PS have an RMSE much higher than the no-noise  and the SKA-noise case (see Fig.~\ref{fig:rmse}). In Fig.~\ref{fig:ps-recon}, we have picked up three randomly selected samples from the test sets, and plotted the predicted power spectra.
We observe that the predictions of the \hi~power spectrum is the best for the no-noise case and for the SKA-noise case. However, sensitivities for HERA is higher than that of SKA, by an order of magnitude around the lowest k-modes. This is reflected in the predictions of the \hi~ signal PS for the test sets corrupted with thermal noise. For the SKA-noise test sets, the reconstruction of the \hi~ 21-cm power spectra is very accurate, and hence the accuracy of the predicted parameters from the second step \texttt{ANN2} is also appreciable, being $\approx80-92\%$. 

\section{Discussions}
\label{discussions}
In this paper, we present a new ANN based machine learning framework to extract the redshifted 21-cm power spectrum and the associated astrophysical parameters from radio observations at relevant low frequencies. To the best of our knowledge, this work is the first demonstration of such an ANN framework on simulated data, in the presence of foreground. We have simulated \hi~ 21-cm power spectra at redshift $z=9.01$, by varying the reionization parameters $\rm{M_{h,min}, \zeta}$ using a semi-numerical code ReionYuga. 

We have considered two separate cases in our work. In the first case, our training set consists of the simulated 21-cm power spectra (PS). We have implemented ANN to extract the reionization parameters from the \hi~ 21-cm power spectrum, and obtained high accuracies for the corresponding test sets, which consists of thermal noise corresponding to 1080 hours of observation of HERA and SKA in addition to the 21-cm PS: $\approx 83-99\%$ for $\rm log~M_{h,min}$ and $\approx 77-94\%$ for $\rm \zeta$. This kind of framework can be extended and applied to datasets from observations in the foreground avoidance regime or for datasets from which the foreground has been very carefully modelled and removed.
In the second case, we have used foreground-dominated datasets for training and have introduced a dual ANN framework to perform the 21-cm power spectrum and parameter extraction. The ANN model from Case 1 is integrated into a two-step ANN framework, to extract the signal power spectrum and then predict the parameters from datasets dominated by foregrounds. The ANN-framework is ignorant about the underlying formulation of how the training sets have been generated. It is trained to extract the \hi~ PS from a dataset that includes foregrounds and eventually learn about the associated reionization parameters.\\
The 21-cm power spectrum extraction was nearly accurate in the noise free case (i.e, case 2a). However, we observed that by adding thermal noise corresponding to two different experiments (SKA and HERA), the accuracy of prediction of the \hi~PS reduced considerably, particularly for HERA. Even a $\lesssim 10\%$ deviation (as shown in Fig.~\ref{fig:ps-recon}) from the original \hi~21-cm PS for HERA, when carried forward to the next stage of the ANN framework (where the astrophysical parameters are predicted), resulted in a much lower R2-score for the predicted parameters as compared to the no-noise case (Case 2a). The reduced accuracy for HERA, from our ANN-framework, could be due to the fact that the \kk-mode sensitivities are much limited for HERA as compared to SKA for a fixed observation time of 1080 hours. \\
The performance of the network would increase if the training datasets included more realistic simulations of the sky (signal+foreground), incorporating detailed instrument models and primary beam information. We are taking into account such realistic synthetic observations along with more sophisticated CD/EoR signal models in an upcoming work. We plan to incorporate these advanced synthetic observations in our future ANN based frameworks.

ANN can be used to extract features from any kind of data by constructing functions which associate the input with the target parameters. In contrast to the existing techniques of parameter estimation (for example, Bayesian frameworks), ANNs do not require a specific, pre-defined prior. The training sets can be thought to be representative of the prior, as used in Bayesian inference. This allows us to use a varied range of shapes of the 21-cm PS as training sets. The use of ML expedites the computational process for parameter estimation considerably. In this work, the training of the networks were done using $1148$ samples. To train an ANN, for example, for the maximum number of epochs ($\sim 5000)$ and batch size corresponding to $\sim 1148$, it took $\sim 5-7$ minutes on a computer with $40$ cores and 62 GB RAM. With smaller batch sizes, it can take up to $\sim20-30$ minutes on the same machine. Use of ANN is computationally very efficient, and can easily predict from the various test sets in less than a minute. In a follow-up work in future, we would like to design a complete signal extraction pipeline which would be trained with several different models of the signal, foreground and will also include the effects of the instrument response, such that it will be a robust tool for predicting various astrophysical parameters associated with the EoR 21-cm signal. 
\section{Acknowledgements}
MC acknowledges the support of DST for providing the INSPIRE fellowship (IF160153).AD acknowledges the support of EMR-II under CSIR.

\section{Data Availability }
Data will be available on request.



\bibliographystyle{mnras}
\bibliography{References} 




\bsp	
\label{lastpage}
\end{document}